\documentclass[twocolumn]{jpsj2} %% two-column layout
\title{Mean-field study of charge order with long periodicity in $\theta$-(BEDT-TTF)$_2$X}% Force line breaks with \\

\author{Masato \textsc{Kaneko}
\thanks{E-mail address: kaneko@hosi.phys.s.u-tokyo.ac.jp} 
and Masao \textsc{Ogata}}

\inst{Department of Physics, University of Tokyo, Tokyo 113-0033}

\abst{
Charge ordering phenomenon in $\theta$-(BEDT-TTF)$_2$X is studied by using 1/4-filled extended Hubbard model on an anisotropic triangular lattice through mean-field approximation. It is found that a metallic charge ordered state with 3-periodicity on lattices (3-fold type charge order) is realized in the realistic parameter region where the nearest neighbor Coulomb interaction $V_{ij}$ is nearly isotropic. This 3-fold state survives up to high temperatures by estimating the free energy. Insulating state with diagonal or vertical-type charge ordering appears as increasing anisotropy of $V_{ij}$. Considering the effect of the structural distortion observed in $\theta$-(BEDT-TTF)$_2$RbZn(SCN)$_4$ at low temperatures, horizontal-type charge ordered insulating phase tends to be stabilized and the region where 3-fold type exists becomes narrower. Our results are consistent with the metallic state with long periodic charge order which can be related to 3-fold type in $\theta$-(BEDT-TTF)$_2$RbZn(SCN)$_4$ at high temperatures and insulating state with horizontal charge order at low temperatures. For $\theta$-(BEDT-TTF)$_2$CsZn(SCN)$_4$, we speculate several kinds of charge-ordered states are energetically competing with each other leading to an inhomogeneous state at low temperatures. }

\kword{charge order, BEDT-TTF, organic conductor, extended Hubbard model, triangular lattice, Hartree approximation}

\begin{document}
\maketitle

\section{\label{sec:1}Introduction}
The family of organic conductors, (BEDT-TTF)$_2$X (in short (ET)$_2$X), which consist of two-dimensional stacking layers of donor BEDT-TTF (ET) molecules has a variety of electronic properties, such as metal-insulator transition, superconductivity and charge ordering, depending on temperature, pressure and the details of its anion molecule X\cite{SHF,H.Mori1}. This indicates that the effect of strong correlation is important and various states are competing with each other. In particular, the value of the nearest neighbor Coulomb interaction is considered to be relatively large in organic compounds and crucial for charge ordering. The various properties observed experimentally have been discussed on the basis of theoretical models which regard the ET molecule as a single site\cite{Kino,Seo,Hotta,SHF,Tanaka,T.Mori,Merino1}.

\begin{figure}
\resizebox{80mm}{!}{\includegraphics{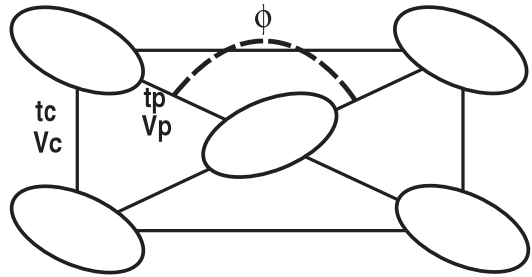}}
\caption{\label{Fig.structure of theta} The structure of $\theta$-type (ET)$_2$X. The two types of transfer integrals and nearest neighbor Coulomb interaction are denoted by $p$ and $c$ for diagonal and vertical direction, respectively. The dihedral angle $\phi$ between molecules is also shown. }
\end{figure}

In this paper, we focus on the $\theta$-type and $\theta_d$-type (ET)$_2$X which have interesting behaviors mentioned later. Here $\theta$ and $\theta_d$ indicate specific lattice structures leading to different transfer integrals and intermolecular interactions. It is considered that the charge ordering phenomena seen in these materials can be theoretically explained in the 1/4-filled extended Hubbard model including nearest neighbor Coulomb repulsion $V$ between ET molecules \cite{Seo}. For the $\theta$-type (ET)$_2$X, the model consists of an anisotropic triangular lattice described in Fig.\ref{Fig.structure of theta}. Two kinds of nearest neighbor Coulomb repulsions and transfer integrals between molecules are denoted by the subscript $p$ and $c$ corresponding to diagonal and vertical direction, respectively. This system is characterized by the dihedral angle $\phi$\cite{H.Mori1} between molecules which depends on pressure or anion X and controls anisotropy expressed as the ratio of nearest neighbor Coulomb repulsion $V_p/V_c$ and transfer integral $t_p/t_c$. According to the angle, various physical properties appear.

For example, $\theta$-(BEDT-TTF)$_{2}$RbZn(SCN)$_4$ shows a metal insulator transition at $T_{\text{MI}}$ = 200K under slow cooling speed. This transition is accompanied by a structural distortion which gives $\theta_d$ structure below $T_{\text{MI}}$. Furthermore, a horizontal stripe charge order is observed. 
In the lower temperature ($T<$ 25K), a spin gap behavior is observed. These properties were clarified based on various experimental measurements \cite{H.Mori1,H.Mori2,Nakamura-Rb,Watanabe1, Miyagawa,Yamamoto,Tajima,Chiba1, Wang1}. However, a recent X-ray scattering experiment\cite{Watanabe1} has shown that, even in the metallic phase, some kinds of short-range charge order with longer periodicity exists. This charge ordering is different from the one seen in the insulating phase below $T_{\text{MI}}$ (i.e., $\theta_d$-phase). Moreover, the NMR studies have shown an anomalous broadening above $T_{\text{MI}}$\cite{Chiba2} and the frequency dependent dielectric constant indicates a feature of insulators\cite{Inagaki2}. In contrast, $\theta$-(BEDT-TTF)$_{2}$CsZn(SCN)$_4$ shows metallic resistivity down to 20K, below which resistivity exhibits sudden increase. A sign of growing horizontal charge order is observed in X-ray measurements \cite{Watanabe2}. Furthermore, large dielectric constant and giant nonlinear conduction are observed at low temperatures, suggesting an appearance of inhomogeneous charge ordered state \cite{Inagaki1}. This is also consistent with NMR and EPR results \cite{Nakamura-Cs, Chiba3} and can be related to the optical conductivity spectra showing a sudden peak at low frequency as lowering temperature \cite{Wang2}.

Having these experiments in mind, we study the 1/4-filled extended Hubbard model in a mean-field approximation. We find that a charge ordered state with long periodicity is realized in $\theta$-phase when the frustration in the nearest-neighbor Coulomb repulsions increases. We call this state as a 3-fold state. Furthermore, we show that the various charge ordered states have similar energies, which will be the origin of the appearance of inhomogeneity observed experimentally. In the $\theta_d$-phase, we show the competition between the 3-fold state and a horizontal stripe state.

The extended Hubbard model has been studied in various contexts. On a square lattice where $V_c$ and $t_c$ are equal to zero, it has been confirmed that a charge ordred state is stabilized above critical values of $V$ \cite{McKenzie,Merino2,Calandra}. However, the actual ratio of $V_p/V_c$ in $\theta$-(ET)$_2$X compounds has been estimated to be comparable, namely $V_p/V_c\simeq1$ \cite{T.Mori}. Thus, it is necessary to investigate an anisotropic triangular lattice where a strong frustration effect due to the lattice structure is expected with respect to the charge degree of freedom\cite{Merino1}. Using realistic transfer integrals, Seo found various patterns of charge order in the ground state by using a mean-field approximation\cite{Seo}. However, he used rather small unit cells and did not take into account of states with longer periodicity. Recently, Mori discussed several states with long periodicity in the atomic limit, i.e., the kinetic energy is completely neglected\cite{T.Mori}. In this paper, we study the unrestricted Hartree approximation in the extended Hubbard model with kinetic energy using the realistic parameters for $\theta$-type and $\theta_d$-type (ET)$_2$X compounds. This enables us to study both metallic and insulating charge ordered state with long periodicity.

In the next section, the model and mean-field approximation is introduced in order to discuss the charge ordering phenomena. The results of calculations which contain various kinds of charge ordered states are presented in $\S$ \ref{sec:3}. The $\theta$-phase and $\theta_d$-phase, realized in $\theta$-(BEDT-TTF)$_2$RbZn(SCN)$_4$ at high and low temperatures, are studied in $\S$ \ref{sec:3-1} and \ref{sec:3-2}, respectively. In $\S$ \ref{sec:4}, we discuss the obtained results in view of the relation to the experimental facts for the actual materials.

\section{\label{sec:2}Method and Formulations}

We study the quarter filled extended Hubbard model on an anisotropic triangular lattice,
\begin{equation}
H=\sum_{<i j>\sigma}(t_{ij}c_i^\dag c_j+h.c)+U\sum_in_{i\uparrow}n_{i\downarrow}+\sum_{<ij>}V_{ij}n_in_j  \label{extended Hubbard}
\end{equation}
where $U$ and $V_{ij}$ are the on-site and the nearest neighbor Coulomb repulsion, respectively, which are crucial for charge order. $V_{ij}$ and $t_{ij}$ depend on the direction between the neighbor molecules, as shown in Fig.\ref{Fig.structure of theta} where the lattice structure of the $\theta$-type (ET)$_2$X and two types of $V_{ij}$ and $t_{ij}$ are shown. We will also study the $\theta_d$-type case later, which has a little complicated $V_{ij}$ and $t_{ij}$ (Fig.\ref{Fig.thetad}). In eq.(\ref{extended Hubbard}), $<ij>$ represents the summation over the pairs of neighbor sites, $\sigma$, $n_{i\sigma}$ and $c_{i\sigma}^{\dag}\;(c_{i\sigma})$ denote up or down spin, number operator ($n_i=n_{i\uparrow}+n_{i\downarrow}$) and the creation (annihilation) operator for an  electron with spin $\sigma$ at $i$th site, respectively. 

The mean-field approximation at zero temperature of the above model was carried out by Seo \cite{Seo}, by decoupling the interaction terms as $n_{i\uparrow}n_{i\downarrow} \rightarrow 
\langle n_{i\uparrow}\rangle n_{i\downarrow}+n_{i\uparrow}\langle n_{i\downarrow}\rangle$.
Although various patterns of charge order could be obtained, Seo studied rather restricted charge order patterns. In order to include the possibility for charge order with long periodicity implied by recent experiments, we carry out an unrestricted Hartree method using the larger number of sites in a unit cell up to 36 sites both at zero and at finite temperatures.

Through mean-field approximation, eq.(\ref{extended Hubbard}) can be written in ${\bf k}$ space as follows,
\begin{equation}
H=\sum_{{\bf k}\sigma}\sum_{l m}\left[\epsilon_{l m {\bf k} \sigma}+\delta_{l m}\biggl(U\langle n_{l {\bf k}-\sigma}\rangle+\sum_{l^{\prime}}V_{l l^{\prime}}\langle n_{l^{\prime}{\bf k}}\rangle\biggr)\right]c_{l {\bf k} \sigma}^{\dag}c_{m {\bf k}\sigma} \label{after MF}
\end{equation}
where $l,l^{\prime}$ and $m$ are the index of sites in a unit cell. By diagonalizing the matrix (\ref{after MF}), the self-consistency equations for the electron density at $i$ th site, $\langle n_{i\sigma}\rangle$ are solved.

To determine which pattern of charge order is the most stable among the various solutions, the ground state energy $E$ and the free energy $F$ at finite temperatures are calculated and compared with each other.
These quantities per unit cell are given by,
\begin{eqnarray}
E_{\text{cell}}&=&\frac{1}{N_{\text{cell}}}\sum_{\bf k}\sum_{l\sigma}E_{l {\bf k} \sigma}n_F(E_{l {\bf k} \sigma})
-\sum_{l}U\langle n_{l\uparrow}\rangle\langle n_{l\downarrow}\rangle \nonumber\\
& &-\sum_{<l m>}V_{l m}\langle n_{l}\rangle\langle n_{m}\rangle, 
\label{energy}\\
F_{\text{cell}}&=&\frac{1}{N_{\text{cell}}}\left[\mu N_{\text{tot}}-\frac{1}{\beta}\sum_{l {\bf k} \sigma}\ln(1+e^{-\beta(E_{l {\bf k} \sigma}-\mu})\right] \nonumber \\
& &-\sum_{l}U\langle n_{l\uparrow}\rangle\langle n_{l\downarrow}\rangle-\sum_{<l m>}V_{l m}\langle n_{l}\rangle\langle n_{m}\rangle, \label{free energy}\\
N_{\text{tot}}&=&\sum_{l m {\bf k}\sigma}|u_{l m {\bf k} \sigma}|^2 n_F(E_{l {\bf k} \sigma}),
\label{total number}
\end{eqnarray}
where $E_{l {\bf k} \sigma}$ and $u_{l m {\bf k} \sigma}$ are the eigenvalue and the element of unitary matrix obtained by diagonalization of (\ref{after MF}), respectively. $n_F(E_{l {\bf k} \sigma})$ is the Fermi distribution function $\frac{1}{1+e^{\beta(E_{l {\bf k}\sigma}-\mu)}}$ and $\mu$ is the chemical potential determined from (\ref{total number}). $N_{\text{tot}}$ and $N_{\text{cell}}$ are the total number of electrons and unit cells, respectively.

\section{\label{sec:3}Results}

\subsection{\label{sec:3-1}$\theta$-type}
First, let us study the case of $\theta$-type (ET)$_2$X as shown in Fig.\ref{Fig.structure of theta}. Coulomb interaction $U$, diagonal hopping $t_p$ and the ratio $V_c/U$ are fixed at $U=0.7eV$, $t_p=0.1eV$ and $V_c/U=0.3$, respectively which are considered as typical values for organic conductors\cite{parameter1,parameter2,parameter3,parameter4}. To investigate the dependence on the anisotropy, the ratio of the transfer integral and nearest neighbor Coulomb interaction are varied as $|t_c/t_p|=0-1$ and $V_p/V_c=0-1$, respectively. This choice of parameters is the same as in ref.\cite{Seo}. As for realistic values, $|t_c/t_p|$ is estimated to be almost equal to 0.1 \cite{transfer,Watanabe1}, while $V_p$ and $V_c$ are comparable. This means that the transfer integrals are similar to the  square lattice, while the nearest neighbor Coulomb interactions have triangular lattice character. In this model, noninteracting energy dispersion is given by $E_{\bf k}= 2t_c\cos(k_y)\pm4t_p\cos(k_x/2)\cos(k_y/2)$. The sign of $t_p$ is irrelevant, while that of $t_c$ is fixed to be positive for the quarter-filled hole system.     

\begin{figure}
\resizebox{80mm}{!}{\includegraphics{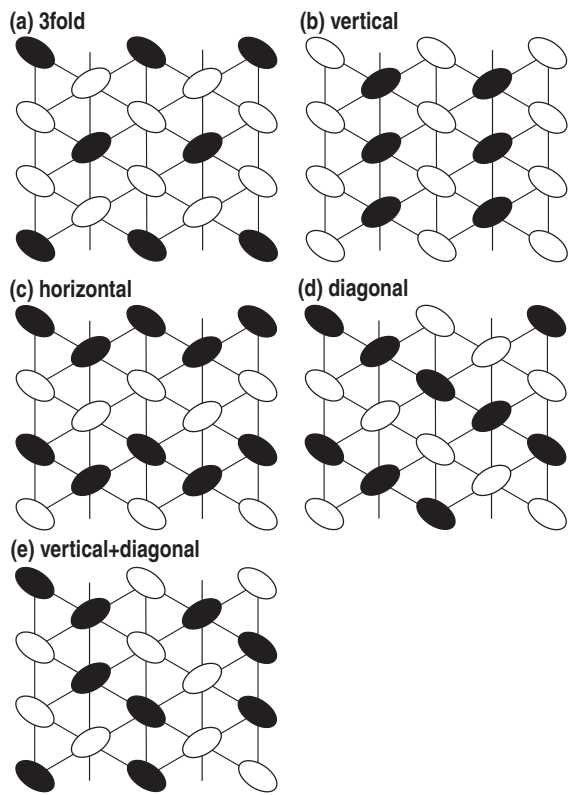}}
\caption{\label{Fig.co} Obtained various patterns of charge order. The charge rich and poor sites are denoted by solid and open ellipses, respectively.}
\end{figure}

We find five charge order patterns as shown in Fig.\ref{Fig.co}, where the charge rich and poor sites as denoted by solid and open ellipsis, respectively. We have tried many other patterns, but they are not stabilized or have a rather higher energy. Fig.\ref{Fig.co}(a) shows a pattern with a unit cell with three lattice sites which was not considered by Seo \cite{Seo}. We call this state as "3-fold" pattern in the following. The other four states are stripe states and we call them as (b)vertical (c)horizontal (d)diagonal (e)vertical+diagonal, respectively.

\begin{figure}
\resizebox{80mm}{!}{\includegraphics{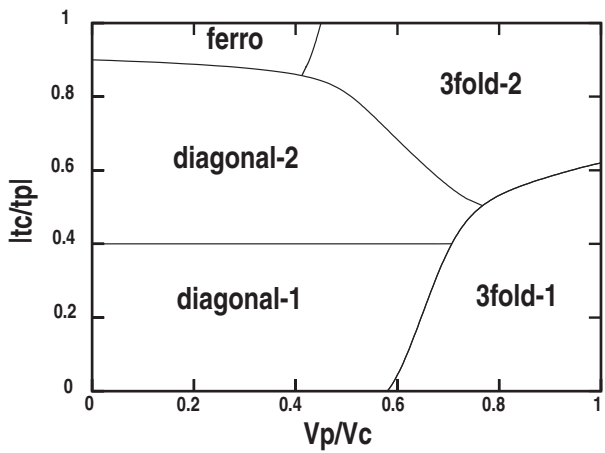}}
\caption{\label{Fig.phasediagram} The phase diagram on $t_c/t_p-V_p/V_c$ plane for $U=0.7eV, V_c/U=0.3, U/t_p=7$. The differences between diagonal-1 and -2 and between 3fold-1 and -2 are spin configuration as explained in Fig.\ref{Fig.co-1}.}
\end{figure}

The obtained phase diagram on $t_c/t_p-V_p/V_c$ plane for $U=0.7eV, V_c/U=0.3, U/t_p=7$ is shown in Fig.\ref{Fig.phasediagram}. Here, the lowest-energy state among the various states is shown. The diagonal-1 and diagonal-2 (and 3-fold-1 and 3-fold-2) have different spin configurations which we discuss shortly. We find that the 3-fold pattern is stabilized in the large $V_p/V_c$ region. The vertical stripe state obtained in \cite{Seo} has a slightly higher energy than the 3-fold pattern in this region. The phase diagram in Fig.\ref{Fig.phasediagram} shows that the frustration in $V_{ij}$ leads to the 3-fold pattern.

\begin{figure}
\resizebox{80mm}{!}{\includegraphics{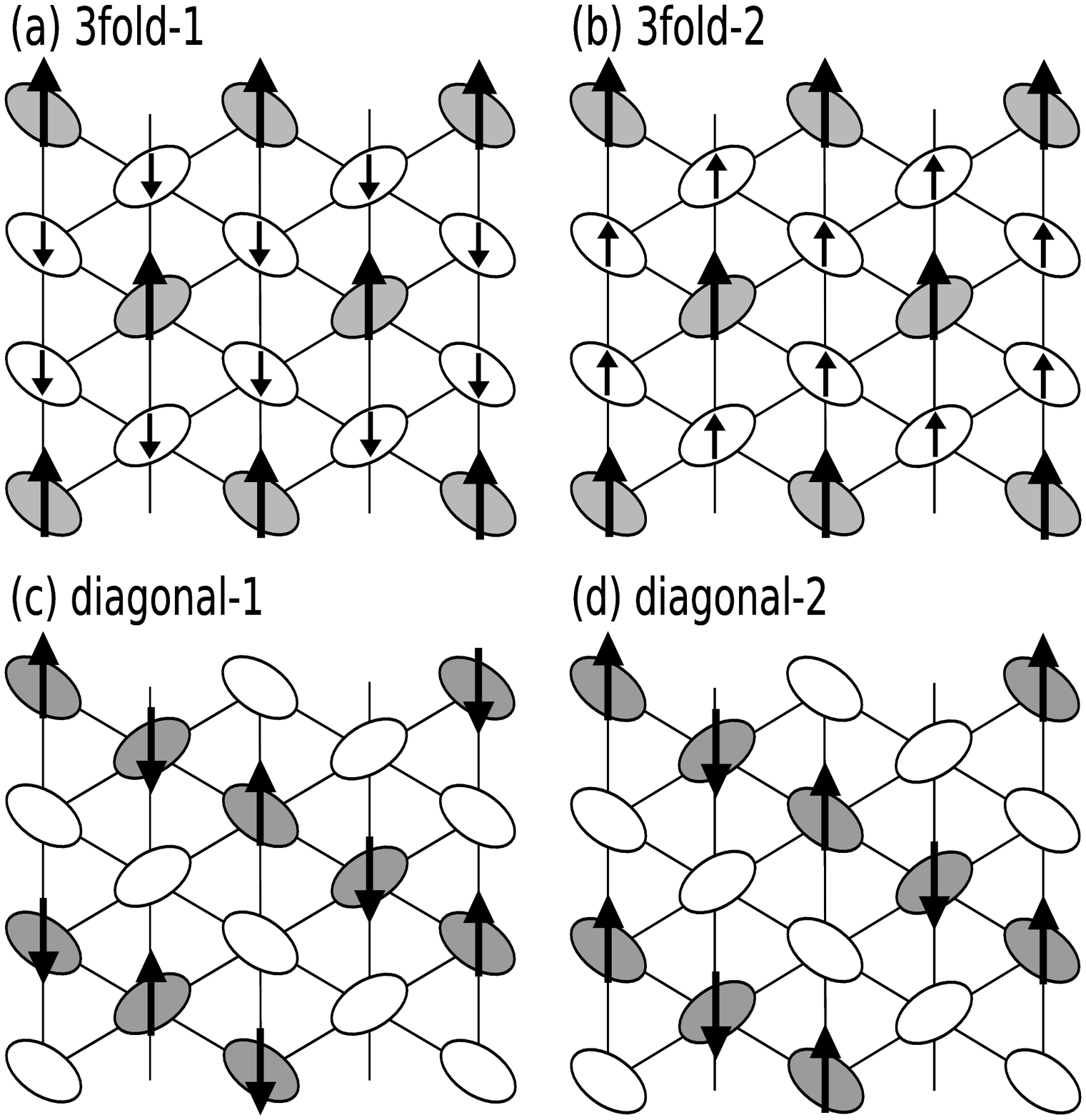}}
\caption{\label{Fig.co-1}3-fold and diagonal type charge order appeared in Fig.\ref{Fig.phasediagram}. The grey and open ellipses represent charge rich and poor sites, respectively. The direction of spin at each site is shown by uparrow or downarrow. The difference between the diagonal-1 and the diagonal-2 is the spin configuration at charge rich sites. The former has antiferromagnetic spin order along $c$ axis between stripes, while the latter has ferromagnetic spin order. 3fold-1 is ferrimagnetic, and 3fold-2 is ferromagnetic.}
\end{figure}  

The hole density at each site depends on the strength of Coulomb interaction and transfer integral. Their values in the ground state range from 0.7 to 0.9 at rich sites and from nearly 0 to 0.1 at poor sites, except for the 3-fold type where they are around 1.0 and 0.25 at rich and poor sites, respectively. 

The diagonal-1 and diagonal-2 states are insulating states in the parameter region where calculations were carried out, since the Fermi energy is located in the band gap. On the other hand, the 3-fold type states are found to be metallic, because the Fermi energy necessarily crosses at least one band among the six bands in the 3-fold type states at quarter-filling.

Next, let us discuss the spin configurations. As shown in Fig.\ref{Fig.co-1}, not only the charge degrees of freedom but also the spin degrees of freedom are ordered. In the stripe states, antiferromagnetic order is obtained along the stripe due to the strong on-site Coulomb interaction leading exchange coupling $J_{ij}$. However, the relative phase of antiferromagnetism between neighboring stripes depends on the inter-chain coupling. The diagonal-1 has antiferromagnetic spin order along the $c$ axis (i.e., vertical direction) between stripes, while the diagonal-2 has ferromagnetic spin order as shown in Fig.\ref{Fig.co-1}(c) and (d). These two states have very close energies and this will be due to the frustration effect of the triangular lattice. On the other hand, in the 3-fold type charge order, the spins on the charge-rich sites have the same direction, as shown in Fig.\ref{Fig.co-1}(a) and (b). The 3-fold-1 has antiferromagnetic spin order between charge rich and poor sites leading to a ferrimagnetic state, while the 3-fold-2 has ferromagnetic spin order. We find that the 3-fold-1 is the most stable in the large $V_p/V_c$ and small $|t_c/t_p|$ region. This is because the gain of the antiferromagnetic exchange energy is the largest among the various spin configurations. In the 3-fold-1, a charge-rich site is surrounded by six charge-poor sites with opposite spin direction.

The phase diagram in Fig.\ref{Fig.phasediagram} shows that the charge order pattern is insensitive to the ratio of transfer $|t_c/t_p|$, although the ordering of spin is varied and ferromagnetic spin configuration is favored as $t_c/t_p$ approaches 1. We think that this tendency to the ferromagnetism is due to the van Hove singularity. When $t_c (>0)$ becomes larger, the van Hove singularity approaches the Fermi energy at quarter filling. This leads to a large density of state which causes the ferromagnetism. For the triangular lattice, it has been shown that the ferromagnetism appears in the large $U$ region irrespective of the electron density\cite{Koretsune}.

\begin{figure}
\resizebox{80mm}{!}{\includegraphics{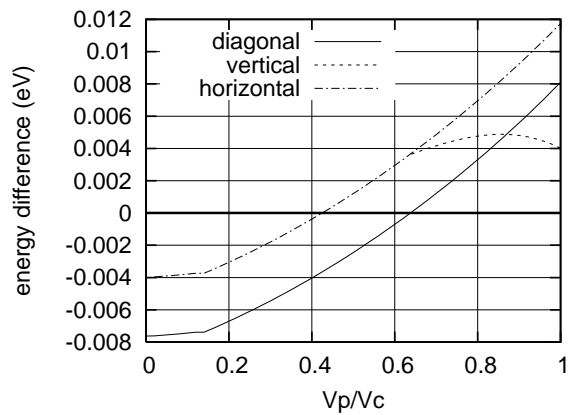}}
\caption{\label{Fig.energydif1} The energy difference between the 3-fold and the other stripe type states for $U=0.7eV, V_c/U=0.3, U/t_p=7$ and $|t_c/t_p|=0.1$. 
The energy of the 3-fold type state is choose to be zero.}
\end{figure}

In Fig.\ref{Fig.energydif1}, we show the ground-state energy difference between the 3-fold and the other stripe-type states for $U=0.7eV, V_c/U=0.3, U/t_p=7$ and $|t_c/t_p|=0.1$. We can see that the 3-fold metallic state is stabilized in the region of $V_p/V_c>0.64$, while insulating state with diagonal stripe charge order is realized for $V_p/V_c<0.64$ where the anisotropic effect by Coulomb interaction is strong. This is reasonable because the ground state with charge order along diagonal is energetically more favorable when $V_p/V_c$ is small enough. In the same way, the vertical type charge order has a lower energy than the diagonal type when $V_p/V_c$ becomes larger. However, the 3-fold type has a much lower energy in this region, as shown in Fig.\ref{Fig.energydif1}.

\begin{figure}
\resizebox{80mm}{!}{\includegraphics{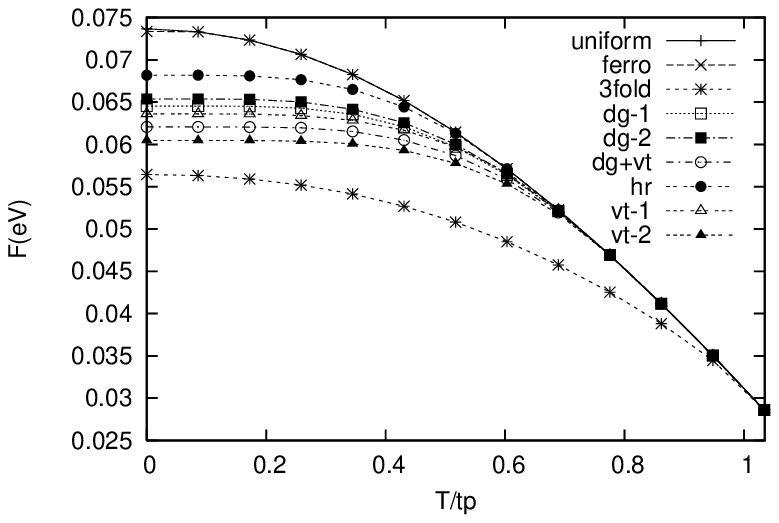}}
\caption{\label{Fig.fe1} Free energy for $V_p/V_c=1$. The other parameters are fixed at $U=0.7eV, V_c/U=0.3, U/t_p=7$ and $|t_c/t_p|=0.1$. 3-fold phase is stable widely for finite temperatures.}
\resizebox{80mm}{!}{\includegraphics{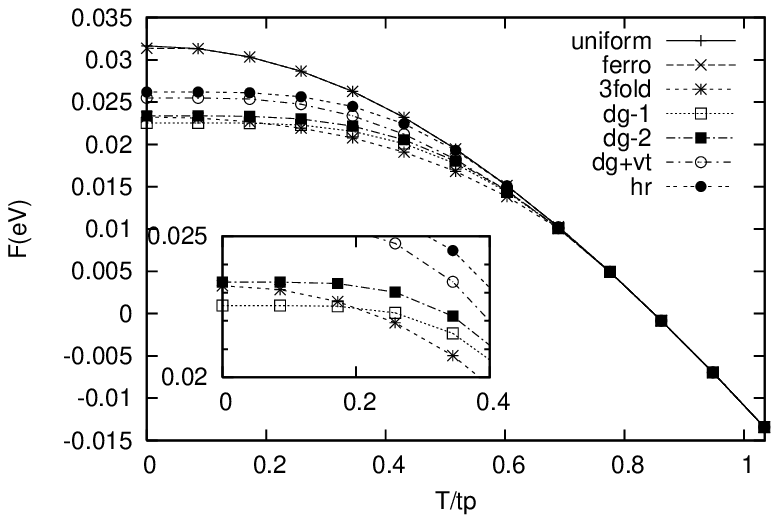}}
\caption{\label{Fig.fe0.6} Free energy for $V_p/V_c=0.6$. The other parameters are fixed at $U=0.7eV, V_c/U=0.3, U/t_p=7$ and $|t_c/t_p|=0.1$. A first-order phase transition between the diagonal-type insulating state and the 3-fold-type metallic state occurs at $T/t_p=0.2$.}
\end{figure}

The behavior at finite temperatures is also investigated by calculating the mean-field free energy F. As seen in Fig.\ref{Fig.fe1}, we find that the 3-fold state gives the lowest free energy in a wide range of temperature when $V_p/V_c\simeq1$. The other charge ordered states have higher free energies which are close to each other. When $V_p/V_c$ is small, the free energy of the state with diagonal charge stripe becomes the lowest. Interestingly, in the intermediate values of $V_p/V_c$, we find that a first-order phase transition between the diagonal stripe state and the 3-fold state at a finite temperature. For example, the free energies at $V_p/V_c=0.6$ are shown in Fig.\ref{Fig.fe0.6}. Apparently the energies of various states become close to each other, and the diagonal-1 state is replaced by the 3-fold state at $T/t_p=0.2$ as the temperature increases. This phase transition shows the higher entropy gain of 3-fold phase represented in the gradient of free energy.

\subsection{\label{sec:3-2}$\theta_d$-type}
\begin{figure}
\resizebox{80mm}{!}{\includegraphics{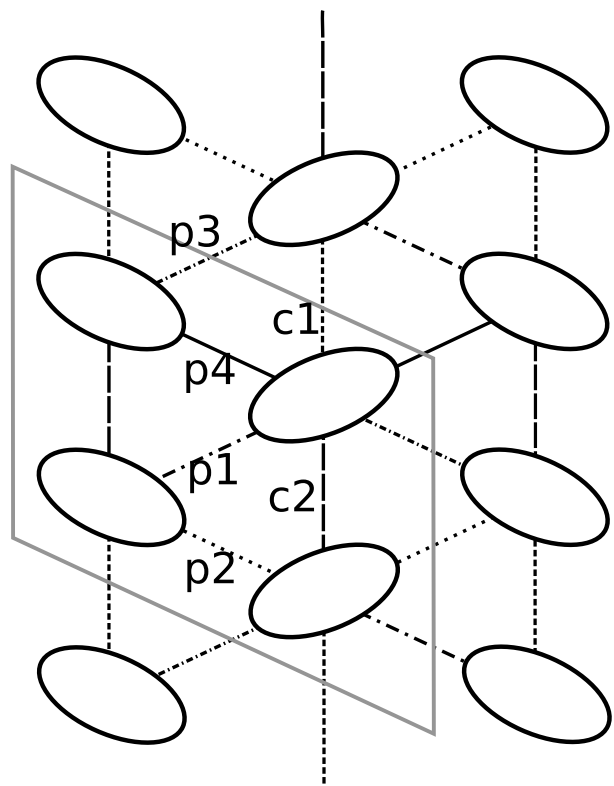}}
\caption{\label{Fig.thetad} $\theta_d$ phase is characterized by the six kinds of transfer integrals $t_{p1}\sim{t_{p4}},\;t_{c1}$ and $t_{c2}$. These values are estimated through extended H\"{u}ckel method as follows, $t_{p1}=16.9,\;t_{p2}=-6.5,\;t_{p3}=2.2,\; t_{p4}=-12.3,\;t_{c1}=1.5 $ and $ t_{c2}=5.2\;(10^{-2}$eV).}
\end{figure}

In this subsection, we discuss the $\theta_d$-phase which is realized in some of (ET$_2$)X materials at low temperatures. The crystal structure of the $\theta_d$-phase is not so simple as the $\theta$-phase due to the lattice distortion. Therefore, different parameters from those of the $\theta$-phase need to be introduced for calculations. In the $\theta_d$-phase, the values of hopping parameter $t_{ij}$ have been estimated through extended H\"{u}ckel method based on experimental data measured on the $\theta$-(BEDT-TTF)RbZn(SCN)$_4$ at $T<T_{\text{MI}}$ \cite{Watanabe1,note_transfer}, which are shown in Fig.\ref{Fig.thetad} together with its structure. The ratio of $V_p/V_c$ is expected to be almost the same as in the $\theta$-phase since the distance between the molecules does not change so much. We carry out calculations by using the set of parameters $t_{ij}$ given in Fig.\ref{Fig.thetad} as varying the ratio $V_p/V_c$ with $U=0.7$eV and $V_c/U=0.3$.

\begin{figure}
\resizebox{80mm}{!}{\includegraphics{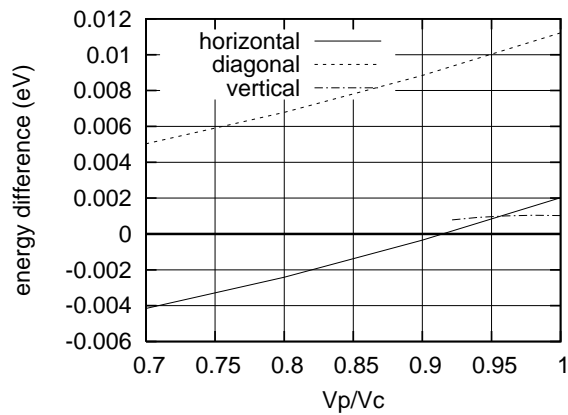}}
\caption{\label{Fig.energydif2} The energy difference between 3-fold, horizontal, vertical and diagonal-type. The energy of the 3-fold type is chosen to be zero. Parameters are fixed at $U=0.7$eV and $V_c/U=0.3$.}
\end{figure}

The energy difference between 3-fold, horizontal, vertical and diagonal-type is shown in Fig.\ref{Fig.energydif2} as a function of $V_p/V_c$. Here, we plot the lowest energy state for each type of charge order among the obtained states with different spin configurations. It is found that, while the 3-fold charge ordered state is stabilized for $V_p/V_c\simeq1$, the horizontal charge ordered state is stabilized for $V_p/V_c<0.92$. Considering the horizontal state is never stabilized in the case of the $\theta$-type, the lattice distortion plays an important role for the horizontal stripe charge order. This tendency is also found in the calculation by exact diagonalization with on-site electron-phonon effects \cite{Clay}.
 
Among the obtained horizontal-type states, the one with antiferromagnetic charge stripe formed along lines connected by the large transfer $t_{p4}$ has the lowest energy. This is due to the effect of exchange coupling between the nearest neighbor spins ($\sim t_{p4}^2/U$). Our result is consistent with the result by Seo where different values of transfer integrals were used. On the other hand, in the 3-fold type realized for the large $V_p/V_c$, there are three patterns of charge distribution in a unit cell with one kind of spin configuration. They are within almost the same energy level and have a tendency with charge rich and poor sites having opposite spin, although this is not so simple as in Fig.\ref{Fig.co-1}(a) and has rather complicated order.

\section{\label{sec:4}Discussions}
In this section, we discuss the results described in the previous section in relation to the charge ordering phenomena observed experimentally in $\theta$-(BEDT-TTF)$_2$X.

The 3-fold type solution obtained in the present calculation can be related to the X-ray scattering experiment\cite{Watanabe1} in the metallic phase above ${T}_{\text{MI}}$ in $\theta$-(BEDT-TTF)$_2$RbZn(SCN)$_4$. From the calculation of the free energy, this 3-fold metallic state is the most stable even at high temperatures in the $\theta$-phase. Our calculation shows that the parameter $V_p/V_c\simeq1$ is important for the stabilization of the 3-fold state. This value is consistent with estimation\cite{T.Mori}. 

However, with respect to the spin degrees of freedom, the lowest-energy state in our approximation is ferrimagnetic, while paramagnetic susceptibility has been observed experimentally\cite{H.Mori1,Nakamura-Rb}. This will be due to the mean-field approximation where the effect of quantum fluctuation is neglected. Thus, in order to discuss the spin degrees of freedom precisely, more appropriate treatment is required, although the charge degrees of freedom are comparable with the actual measurements.

Below the metal-insulator transition, $\theta$-(BEDT-TTF)$_2$RbZn(SCN)$_4$ changes its structure from $\theta$ to $\theta_d$-type. For this $\theta_d$ case, we find the stabilization of the horizontal stripe which is consistent with experiments\cite{Watanabe1,Chiba1,Tajima,Miyagawa,Yamamoto,Wang1}. The horizontal stripe charge order is observed not only in X$=$RbM$^{\prime}$ but also in X$=$TlM$^{\prime}$ (M$^{\prime}$ = Zn, Co) in the $\theta_d$-phase\cite{Suzuki}. Our result suggests that, in addition to the ratio $V_p/V_c$, the network of transfer integral $t_{ij}$ plays an important role for the stabilization of the charge order patterns. The magnetic susceptibility by experiments can be fit for the the Bonner-Fisher curve for the one-dimensional Heisenberg model \cite{H.Mori1}. Although the mean-field calculation gives an antiferromagnetic spin order along the horizontal stripe, it is natural to have a one-dimensional Heisenberg spin fluctuation at finite temperatures when the quantum fluctuations are taken into account.

In the present paper, we did not discuss the phase transition between $\theta$ and $\theta_d$-type structure observed in $\theta$-(ET)$_2$RbZn(SCN)$_4$. For this purpose, the elastic energies, or lattice distortion energies should be taken into account, which is beyond the scope of the present paper. Generally, the symmetry in the $\theta$-phase can be broken to a lower-symmetry as the temperature decreases. Thus, it is natural to expect $\theta_d$-phase with insulating horizontal stripe at low temperatures.
 
Experimentally, the other type of charge ordered states, i.e., diagonal and vertical-type have been actually found. For example, $\theta_m$-TlZn exhibits the former type \cite{Suzuki}. Here, $\theta_m$ represents a monoclinic structure in contrast to an orthorhombic one in $\theta$ and $\theta_d$-phase. On the other hand, $\theta$-(BDT-TTP)$_2$Cu(NCS)$_2$ having similar structure to the $\theta$-type exhibits a checkerboard type charge order \cite{Yakushi}. Thus, the implication of various competing states makes the system complicated and causes interesting physical properties. 
 
In the case of $\theta$-(BEDT-TTF)$_2$CsZn(SCN)$_4$, the inhomogeneous emergence of charge ordered insulating domains is suggested \cite{Inagaki1}. These domains seem to coexist with other metallic domains. On the other hand, the X-ray diffraction study shows that a horizontal charge order grows blow 50K \cite{Watanabe2}. These experimental observations will be related to our result where 3-fold metallic state and horizontal insulating charge ordered state are competing with each other energetically. In the phase diagram for $\theta$-(ET)$_2$X families, $\theta$-(ET)$_2$CsZn(SCN)$_4$ is located near the boundary between  $\theta$ and the $\theta_d$-structure\cite{H.Mori1}. Therefore, it is possible that the strength of coupling to the lattice grows only in some local domains at low temperatures. As a result, competing phases can gradually form in each domain, although the global structural phase transition to $\theta_d$-type as in $\theta$-(ET)$_2$RbZn(SCN)$_4$ does not occur.

In summary, we investigated the charge ordering phenomenon in $\theta$-(ET)$_2$X by using 1/4-filled extended Hubbard model on an anisotropic triangular lattice through mean-field approximation. It is found that metallic state with 3-fold type charge order is realized in the $\theta$-phase in the parameter region where the nearest neighbor Coulomb interaction $V_{ij}$ is nearly isotropic. This 3-fold state survives up to high temperatures. Insulating state with diagonal or vertical-type charge ordering appears as increasing anisotropy of $V_{ij}$. On the other hand, in the $\theta_d$-phase, horizontal-type charge ordered insulating phase is stabilized and the region where 3-fold type exists becomes narrower.

These results are consistent with the metallic state with long periodic charge order which can be related to 3-fold type in $\theta$-(ET)$_2$RbZn(SCN)$_4$ at high temperatures. The insulating state with horizontal charge order is enhanced in the $\theta_d$-phase at low temperatures. For $\theta$-(ET)$_2$CsZn(SCN)$_4$, this effect is considered to play an important role within local domains, leading to the interesting competition with metallic state at low temperatures.  

\section*{\label{sec:final}Acknowledgements}
We thank Y.Tanaka and H.Watanabe for valuable discussions.


\begin{thebibliography}{99}
\bibitem{note_transfer} Note that there have been some confusions in literatures about the signs of $t_{ij}$. Our choice of $t_{ij}$ gives the reasonable Fermi surface.
\end{thebibliography}
\end{document}